# Collective flow and balance energy in asymmetric heavy-ion collisions

Supriya Goyal, S. Gautam, Aman D. Sood, and Rajeev K. Puri,*
*Department of Physics, Panjab University, Chandigarh-160014, INDIA*
*\*email: rkpuri@pu.ac.in*

## Introduction

Heavy-ion collisions at intermediate energy provide a useful tool to determine the nuclear matter equation of state as well as in-medium nucleon-nucleon (nn) cross-section. One of the most promising phenomena that can be helpful in this context is the *disappearance of transverse in-plane flow*. At low incident energies, the dominance of attractive interactions results in the emission of particles into backward hemisphere which leads to negative transverse in-plane flow whereas at higher incident energies, the particles are emitted into forward hemisphere resulting in positive flow. As one goes from low incident energy to higher one there exist a particular energy at which the net transverse in-plane flow disappears. This energy is termed as the *energy of vanishing flow* (EVF) [1, 2]. Up to now, one has measured the energy of vanishing flow in the reactions of $^{12}C+^{12}C$, $^{20}Ne+^{27}Al$, $^{36}Ar+^{27}Al$, $^{40}Ar+^{27}Al$, $^{40}Ar+^{45}Sc$, $^{40}Ar+^{51}V$, $^{64}Zn+^{27}Al$, $^{40}Ar+^{58}Ni$, $^{64}Zn+^{48}Ti$, $^{58}Ni+^{58}Ni$, $^{58}Fe+^{58}Fe$, $^{64}Zn+^{58}Ni$, $^{86}Kr+^{93}Nb$, $^{93}Nb+^{93}Nb$, $^{129}Xe+^{118}Sn$, $^{139}La+^{139}La$, and $^{197}Au+^{197}Au$ [3]. However most of the reactions studied are nearly symmetric. Since, the dynamics of heavy-ion reactions also depends on asymmetry of the reaction [4]. We therefore, aim to study the transverse in-plane flow as well as its disappearance for different colliding nuclei with varying asymmetry. We plan to address this question using quantum molecular dynamics (QMD) model [5].

## Model

The QMD model is an n-body theory that simulates the heavy-ion reactions on event by event basis. This is based on a molecular dynamics picture where nucleons interact via two and three-body interactions. The nucleons propagate according to the classical equations of motion:

$$\frac{d\mathbf{p}_i}{dt} = -\frac{dH}{d\mathbf{r}_i} \quad and \quad \frac{d\mathbf{r}_i}{dt} = \frac{dH}{d\mathbf{p}_i}, \quad (1)$$

where H stands for the Hamiltonian which is given by

$$H = \sum_i \left[ T_i + \frac{1}{2} \sum_{ij} V_{ij} \right]. \quad (2)$$

$T_i$ is the kinetic energy term and $V_{ij}$ is the nuclear potential which consists of

$$V_{ij} = V^{Skyrme} + V^{Yuk} + V^{Coul}, \quad (3)$$

where $V^{Skyrme}$, $V^{Yuk}$, and $V^{Coul}$ are, respectively, the local Skyrme, Yukawa, and Coulomb potentials. The final transverse in-plane flow is calculated using

$$\left\langle p_x^{dir} \right\rangle = \frac{1}{A} \sum_{i=1}^{A} sign\{y(i)\} \mathbf{p_x}(i). \quad (4)$$

Here $y(i)$ is the rapidity and $\mathbf{p_x}(i)$ is the transverse momentum of $i^{th}$ particle. The rapidity is defined as:

$$y(i) = \frac{1}{2} \ln \frac{\mathbf{E}(i) + \mathbf{p}_z(i)}{\mathbf{E}(i) - \mathbf{p}_z(i)}, \quad (5)$$

where $\mathbf{E}(i)$ and $\mathbf{p}_z(i)$ are, respectively, the total energy and longitudinal momentum of $i^{th}$ particle. The EVF was then deduced using a straight line interpolation.

## Results and discussion

Here we study the collisions of various projectiles and targets with varying degree of asymmetry parameter ($\eta$) keeping the combined mass of the system fixed (100 in our case), where $\eta$ is defined as

$$\eta = \left|\frac{A_T - A_P}{A_T + A_P}\right|. \qquad (6)$$

$A_T$ and $A_P$ are the masses of target and projectile, respectively. For symmetric collisions $\eta = 0$ and nonzero values of $\eta$ corresponds to different asymmetries. We simulated the collisions with different $\eta$ ranging from 0-0.86 at different incident energies varying from 40 to 400 MeV/nucleon in small steps and at reduced impact parameter (b/$b_{max}$) of 0.2. For the present study, we employed a stiff equation of state (K=380 MeV) along with a constant isotropic cross-section of 40 mb strength. In fig. 1, we show $<p_x^{dir}>$ for collisions with different asymmetry as a function of the incident energy. The lines are to guide the eye. The transverse in-plane flow increases monotonically with increase in incident energy in all the cases, which is due to the increase in number of nucleon-nucleon collisions as well as dominance of repulsive mean field. At fixed incident energy, $<p_x^{dir}>$ decreases as the asymmetry of reaction increases in the high energy region whereas the trend gets reversed at very low incident energies (e.g. 40 MeV/nucleon). The energy of vanishing flow remains nearly same up to $\eta = 0.52$ after which it increases sharply which may be due to decrease in the number of collisions as well as Coulomb repulsion.

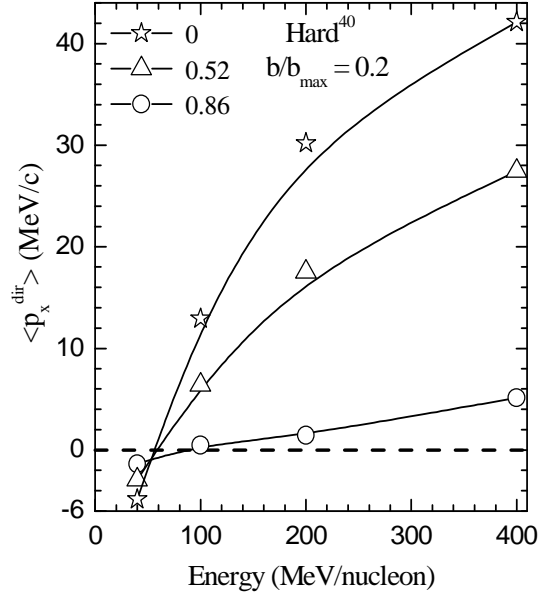

**Fig. 1** The transverse in-plane flow $<p_x^{dir}>$ as a function of incident energy for different aymmetric systems having same mass.